\documentclass[aps, pre,amsmath,amssymb,twocolumn,floatfix,eqsecnum]{revtex4}

\newif\ifpdf\ifx\pdfoutput\undefined\pdffalse\else\pdfoutput=1\pdftrue\fi

\usepackage{graphicx}
\usepackage{bm}
\usepackage{epsfig}

\newcommand{\be}{\begin{equation}}

\newcommand{\ee}{\end{equation}}

\begin{document}

\title{\bf Liquid-vapor interface of a polydisperse fluid}

\author{Matteo Buzzacchi}
\author{Nigel B. Wilding}
\affiliation{Department of Physics, University of Bath, Bath BA2 7AY, United Kingdom}

\date{\today}

\begin{abstract} 

We report a Grand Canonical Monte Carlo simulation study of the
liquid-vapor interface of a model fluid exhibiting polydispersity in
terms of the particle size $\sigma$. The bulk density distribution,
$\rho^0(\sigma)$, of the system is controlled by the imposed chemical
potential distribution $\mu(\sigma)$. We choose the latter such that
$\rho^0(\sigma)$ assumes a Schulz form with associated degree of
polydispersity $\approx 14\%$. By introducing a smooth attractive wall,
a planar liquid-vapor interface is formed for bulk state points within
the region of liquid-vapor coexistence. Owing to fractionation, the
pure liquid phase is enriched in large particles, with respect to the
coexisting vapor. We investigate how the spatial non-uniformity of the
density near the liquid-vapor interface affects the evolution of the
local distribution of particle sizes between the limiting pure phase
forms. We find (as previously predicted by density functional theory,
Bellier-Castella {\em et al}, Phys. Rev. {\bf E65}, 021503 (2002)) a
segregation of smaller particles to the interface. The magnitude of
this effect is quantified for various $\sigma$ via measurements of the
relative adsorption. Additionally, we consider the utility of various
estimators for the interfacial width and highlight the difficulties of
isolating the intrinsic contribution of polydispersity to this width.


\end{abstract} 
\maketitle
\setcounter{totalnumber}{10}

\section{Introduction and background}

\label{sec:intro}

Complex fluids in which the particles are similar in character but not
strictly identical, are termed polydisperse. Examples of such arise
throughout soft matter science, notably in colloidal dispersions,
polymer solutions and liquid-crystals. Typically the polydispersity of
these systems is manifest as variation in some physical attribute such
as size, shape or charge, which one denotes by a continuous parameter
$\sigma$. The form of the polydispersity is then quantifiable in terms
of a density distribution $\rho(\sigma)$ measuring the number density
of particles of each $\sigma$. Accordingly one can regard the system as
a mixture of an infinite number of particle ``species'' each labelled
by the value of $\sigma$ \cite{SALACUSE}.

As has long been appreciated, polydispersity can deeply influence the
thermodynamical and processing properties of complex fluids
\cite{LARSON99,CHAIKIN00,CHANDRASEKHAR92,ELIAS97}, making a clear
elucidation of its detailed role a matter of both fundamental and
practical importance. The majority of recent effort in this regard has
focussed on clarifying the bulk phase behaviour of polydisperse systems
(see \cite{SOLLICH02} for a recent review), which is known to be
considerable richer in both variety and character than that of
corresponding monodisperse systems. The source of this richness can be
traced to {\em fractionation} effects: at coexistence a polydisperse
fluid described by some initial ``parent'' distribution,
$\rho^0(\sigma)$,  may split into two or more `daughter' phases
$\rho^{(\alpha)}(\sigma)$, $\alpha=1,2,...$, each of which differs in
composition from the parent. The sole constraint is that the 
volumetric average of the daughter distributions equals the parent
distribution.  

\begin{figure}[h]
\includegraphics[width=6.5cm,clip=true]{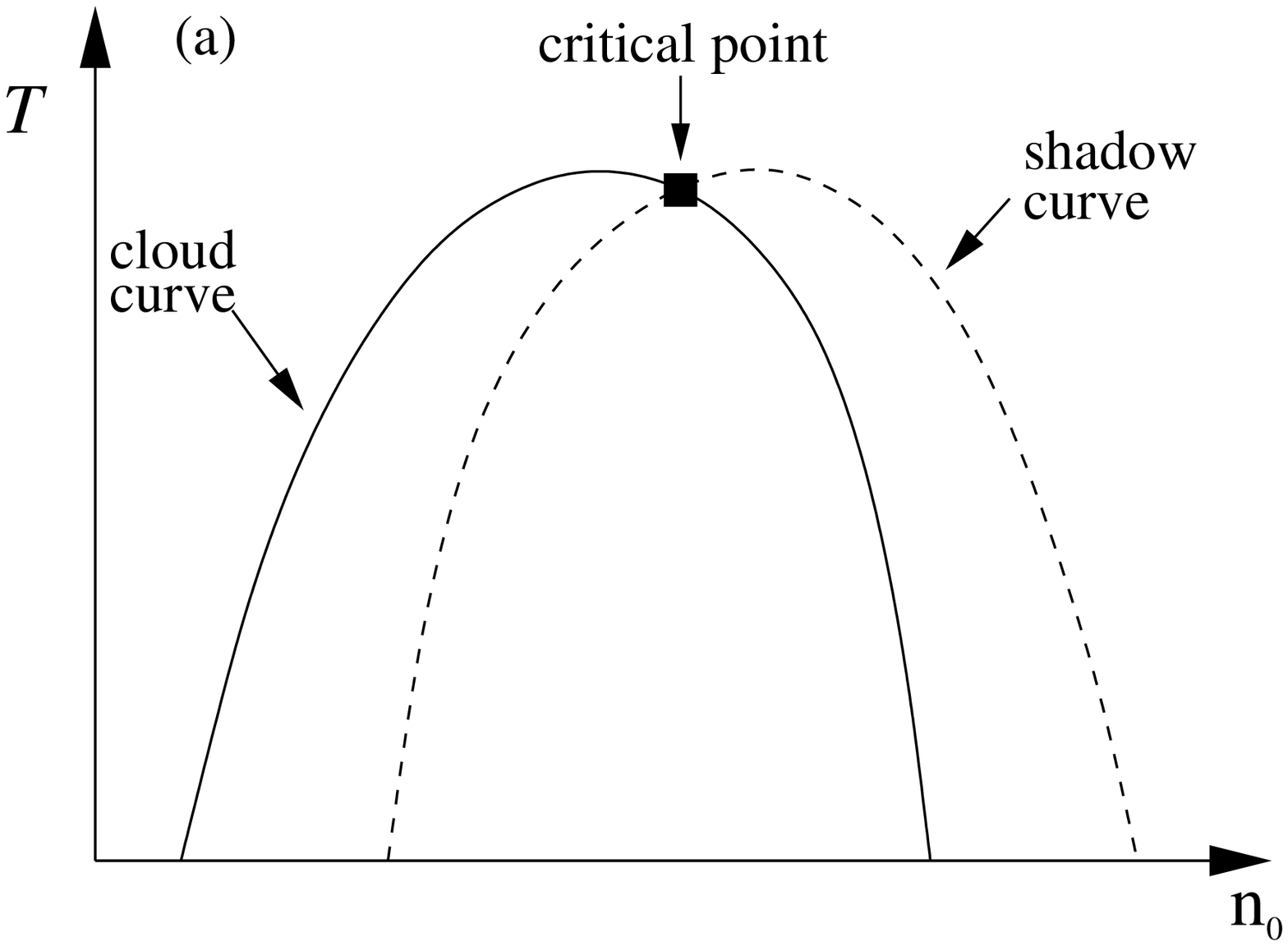}
\includegraphics[width=6.5cm,clip=true]{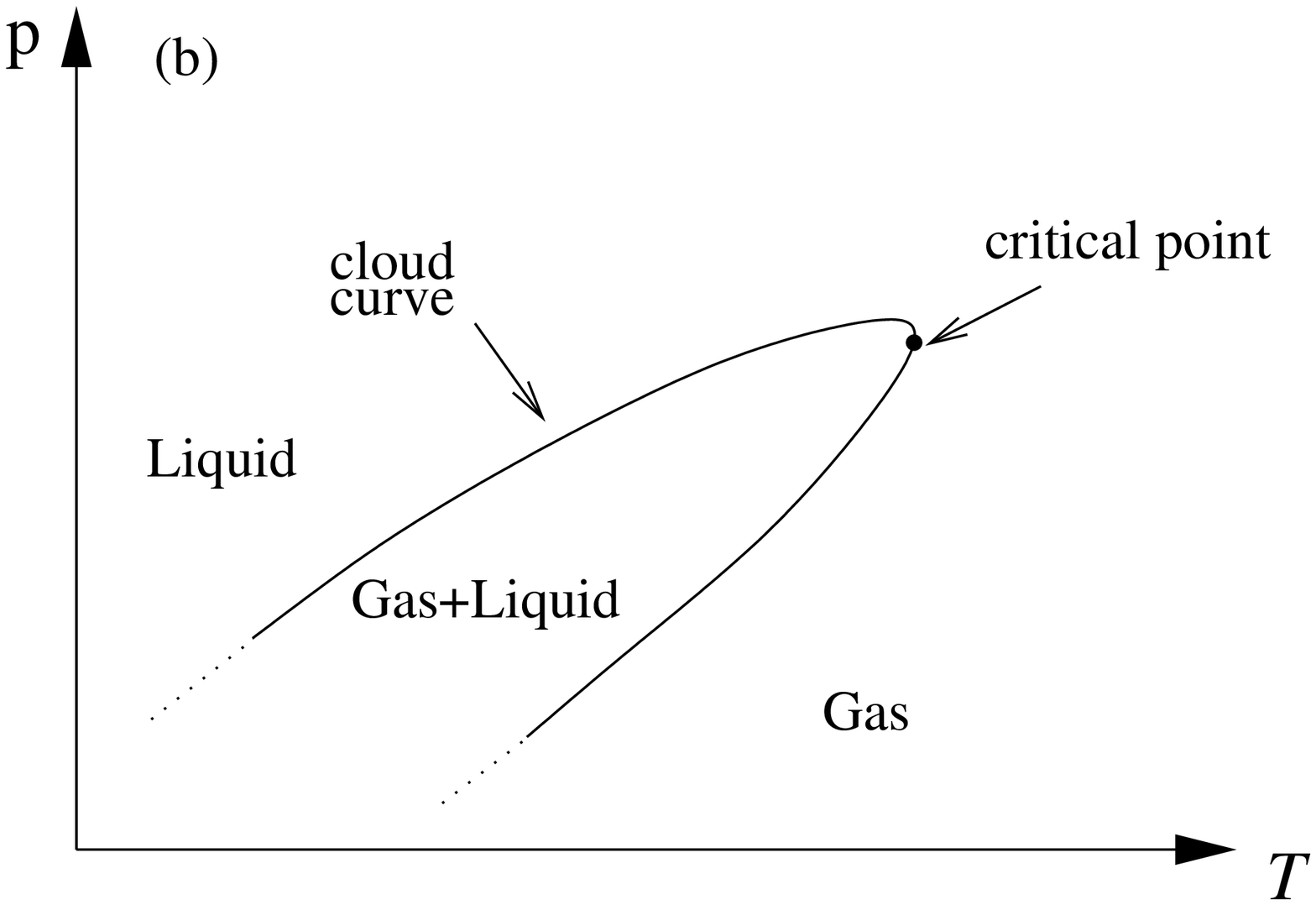}
\caption{Schematic phase diagram as described in the text. {\bf (a)} The $n_0$-$T$
plane. {\bf (b)} The $p$-$T$ plane.}
\label{fig:schem}
\end{figure}

The effect of fractionation on phase diagrams can be dramatic. For
instance, the familiar liquid-vapor binodal in the density-temperature
plane of a monodisperse fluid splits into cloud and shadow curves
\cite{SOLLICH02}, as shown schematically in fig.~\ref{fig:schem}(a).
These mark, respectively, the density of the onset of phase separation
and the density of the incipient (shadow) phase. The critical point
occurs neither at the extremum of the cloud nor the shadow curve but at
their intersection. One implication of this is that even at the
critical temperature, liquid vapor coexistence can occur provided the
overall parent density is less than its critical value.  Additional
insight into fractionation effects can be gleaned from the
pressure-temperature plane of the phase diagram. In a monodisperse
system, coexistence occurs along a line in this plane which terminates
at the critical point. However, the introduction of polydispersity
broadens this line into a {\em region} having a `banana'-like shape
\cite{RASCON03,BELLIER00}. The critical point generally lies neither at
a point of maximum temperature nor maximum pressure on the perimeter of
this region (fig.~\ref{fig:schem}(b)). Traversing the coexistence
region from one pure phase to the other (eg. along an isobar or
isotherm) corresponds to {\em smoothly} varying the relative volumes of the
system occupied by the two coexisting phases, with concomitant smooth
variation in the forms of the daughter distributions.

In view of the richness of the bulk phase behaviour, it is natural to
enquire how polydispersity influences fluid interfacial properties.
Previous work on this subject has largely been limited to the study of
size-disperse hard spheres either at a single hard wall
\cite{BRYK,PAGONA00,BUZZ04,KEGEL04} or confined to a planar slit
\cite{PIZIO01,KIM01}. These investigations focussed principally on the
effect of the wall on the local fluid structure and distribution of
particle sizes. However, since no explicit interparticle attraction was
considered, issues of wetting and fluid-fluid coexistence did not
arise. The sole previous study of a fluid-fluid interface in a
polydisperse system (of which we are aware) is the density functional
theory study of Bellier-Castella {\em et al} \cite{BELLIER02}. Building
on previous studies of a homogeneous polydisperse van der Waals fluid
\cite{BELLIER00}, these authors obtained the density profiles for
various species across a planar liquid-vapor interface. Their
calculations indicated a preferential adsorption of small particles at
the interface which they reported to be broadened with respect to the
monodisperse limit. 

In the present work, we have attempted to extend the understanding of
interfacial behaviour in polydisperse fluids by performing detailed
Monte Carlo simulations of a system of spherical size-disperse
particles interacting via a Lennard-Jones potential. The simulations
are performed within the grand canonical ensemble and employ a chemical
potential distribution $\mu(\sigma)$, the form of which is chosen such
as to yield a bulk density distribution $\rho^0(\sigma)$ having a fixed
Schulz form, with associated degree of polydispersity $\approx 14\%$.
Such ``fixed polydispersity'' corresponds to the experimental situation
in, for example, colloidal dispersions or polymer solutions, where the
form of the parent distribution is prescribed by the synthesis process
of the particles, and only its scale can change depending on the
quantity of solvent present.

\section{Method}

\subsection{Model and observables}
\label{sec:model}

The model we consider comprises a system of particles interacting via an
interparticle potential of the Lennard-Jones (LJ) form:

\begin{equation}
u_{ij}=4\epsilon_{ij}\left[\left(\frac{\sigma_{ij}}{r_{ij}}\right)^{12}-\left(\frac{\sigma_{ij}}{r_{ij}}\right)^6\right]\:.
\label{eq:ljpot}
\end{equation}
Here $r_{ij}=|\bf{r}_i-\bf{r}_j|$ is the particle separation and we
employ the mixing rules $\sigma_{ij}=(\sigma_i+\sigma_j)/2$ and
$\epsilon_{ij}=\epsilon\sigma_i\sigma_j$.   A cutoff was applied to the
potential for particle separations $r_{ij}>2.5\sigma_{ij}$.

Conventionally the state of such a fluid in the bulk is described by a
parent density distribution $\rho^0(\sigma)$, which
can be written \cite{SALACUSE}:

\begin{equation}
\rho^0(\sigma)=n_0f(\sigma)\:.
\label{eq:parent}
\end{equation}
Here $n_0=N/V$ is the overall particle number density, while
$f(\sigma)$ is a normalized shape function whose average value
$\bar\sigma$ serves to set the scale for all lengths. Since the form of
$f(\sigma)$ is fixed, the bulk phase diagram is spanned by $n_0$ and
the temperature $T$ (cf. fig~\ref{fig:schem}(a)). 

\begin{figure}[h]
\includegraphics[width=6.5cm,clip=true]{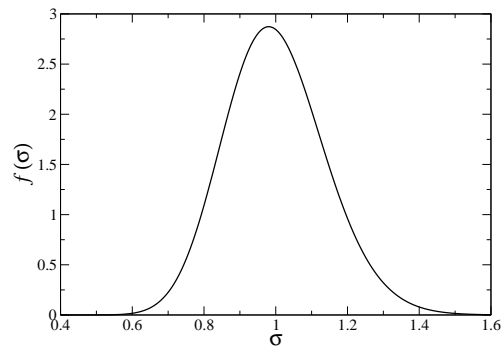}
\caption{The Schulz parent size distribution $f(\sigma)$ studied in this
work.}
\label{fig:schulz}
\end{figure}

A commonly used measure of the scale of variation in the particle
diameters is provided by the dimensionless degree of polydispersity,
defined as the standard deviation of the parent distribution,
normalized by its mean:


\begin{equation}
\delta=\frac{\sqrt{\overline{(\sigma-\bar{\sigma})^2}}}{\bar\sigma}.
\label{eq:poly}
\end{equation}
In the present work, we have elected to study a parent of the Schulz form:

\begin{equation}
f(\sigma)=\frac{1}{Z!}\left(\frac{Z+1}{\bar{\sigma}}\right)^{Z+1}\sigma^Z\exp\left[-\left(\frac{Z+1}{\bar{\sigma}}\right)\sigma\right]\:.
\label{eq:schulz}
\end{equation}
Here the parameter $Z$ controls the width of the distribution and thence
the value of $\delta$. We have considered the case $Z=50$,
corresponding to $\delta=(Z+1)^{-1/2}\simeq 0.14$. The resulting form of
$f(\sigma)$ is shown in fig.~\ref{fig:schulz}. Note that in
contrast to, for example, a Gaussian, the Schulz distribution vanishes smoothly 
(has a natural cutoff) as $\sigma\to 0$. For the purposes of the MC
simulations described below, however, one does require an upper cutoff
in $\sigma$, beyond which $f(\sigma)$ is truncated. We set this to be
$\sigma_c=1.6$. 

In general, a polydisperse fluid described by eqs.~\ref{eq:ljpot} will
exhibit liquid-vapor coexistence within the region of the phase
diagram enclosed by the cloud curve (cf. fig.~\ref{fig:schem}).  For
such coexistence states, a liquid-vapor interface can be formed by
introducing a sufficiently attractive wall in the plane $z=0$. In this
work, the particle-wall interactions are assigned the form

\begin{equation}
u_{pw}(\sigma_i,z_i)=\epsilon_{w}\sigma_i\bar\sigma\left[ \frac{2}{5}\left(
\frac{\sigma_{iw}}{z_{i}} \right)^{10}
-\left( \frac{\sigma_{iw}}{z_{i}} \right)^{4} \right],
\label{eq:pw_pot}
\end{equation}
where $\sigma_{iw}=(\sigma_i+\bar\sigma)/2$. The form of this potential
derives from regarding the wall as a mono-layer of monodisperse
particles having diameter $\bar\sigma$ and interacting with fluid
particles via eq.~\ref{eq:ljpot}. 

In order to quantify the resulting interfacial properties, we consider the
ensemble averaged local density distribution $\rho(\sigma,z)$ at a
perpendicular distance $z$ from the wall:

\begin{equation}
\rho(\sigma,z)=\int_0^{L_x}  \int_0^{L_y} \rho(\sigma,{\bf r})\; dx dy\:.
\end{equation}
We shall also find it useful to define a number of one-dimensional profiles which
derive from $\rho(\sigma,z)$. These are the overall density profile

\begin{equation}
\rho(z)=\int~d\sigma~\rho(\sigma,z)\:;
\label{eq:denprof}
\end{equation} 
the volume fraction profile

\begin{equation}
\eta(z)=\frac{\pi}{6}\int~d\sigma~\rho(\sigma,z)\sigma^3\:;
\label{eq:vfracprof}
\end{equation} 
the local concentration profile:

\begin{equation}
\phi(\sigma,z)=\frac{\rho(\sigma,z)}{\rho(z)}\:;
\label{eq:phidef}
\end{equation} 
and finally, (in analogy with eq.~\ref{eq:parent}), the local normalized size distribution:

\begin{equation}
f(\sigma|z)=\frac{\rho(\sigma|z)}{\int d\sigma \rho(\sigma|z)}\:.
\label{eq:locszdist}
\end{equation}

The utility of all these quantities will become apparent in sec.~\ref{sec:results}.

\subsection{Simulation methodology}

\label{sec:method}

The grand canonical Monte Carlo algorithm utilized to study our model
deploys four types of operation: particle displacements, deletions,
insertions, and resizing. The particle diameter $\sigma$ is treated as
a continuous variable throughout. Observables, on the other hand (principally the
form of $\rho(\sigma,z)$) are accumulated as histograms, formed by
discretising into bins the permitted ranges of $\sigma<\sigma_c$ (see
sec.~\ref{sec:model}) and $z$ (i.e. $0<z<L_z$). The bin widths used
were $\delta\sigma=0.02$ and $\delta z=0.02$. Further details of the
implementation can be found elsewhere~\cite{WILDING02}.

The algorithm requires as input a chemical potential distribution
$\mu(\sigma)$. Given a nominated shape function $f(\sigma)$,
specification of the overall density $n_0$ and the temperature $T$
serves to fix the bulk form of $\mu(\sigma|n_0,T)$. The assumption that
the confined fluid exists in equilibrium with a bulk reservoir implies
that both the bulk and confined systems have equal $\mu(\sigma)$. Thus
in order to perform simulations of the fluid at a wall, we first
require the form of $\mu(\sigma)$ at specified points inside the bulk
coexistence region. For the present work, we have focused on the
critical isotherm $\tilde T_c=k_BT_c/\epsilon=1.384$, the critical
temperature having been determined in a separate study
\cite{WILDING05}. Recall (sec.~\ref{sec:intro}) that liquid-vapor
coexistence can occur in a polydisperse fluid even at the critical
temperature, provided that the overall density $n_0$ lies between the
densities of the cloud point and critical point. At $\tilde T_c$, the
cloud point occurs at $n_0=0.044(1)$ and the critical point at
$n_0=0.326(2)$.

Using fully periodic simulations, we have determined the form of
$\mu(\sigma)$ for a selection of values of $n_0$ along the critical
isotherm between the cloud point and the critical point. For this
purpose, the combined techniques of non-equilibrium potential
refinement and histogram reweighting were utilized. A description of
the procedure have been previously presented in
refs.~\cite{WILDING03,WILDING04A,WILDING04B} and we again refer the
interested reader to these papers for full details.

The resulting forms of $\mu(\sigma|n_0,\tilde T_c)$ for the respective
values of $n_0$ were then used to study the effects of introducing two
oppositely facing walls at $z=0$ and $z=L_z$ (the system remaining
periodic in the $x$ and $y$ directions).  The system size was set at
$L_x=L_y=15\bar\sigma, L_z=40\bar\sigma$. Interactions between the particles and the
wall at $z=0$ were assigned the form eq.~\ref{eq:pw_pot}, while a
simple hard (impenetrable) wall condition was applied at $z=L_z$. This
arrangement ensures that only the wall at $z=0$ can become wet. 

For values of $n_0$ near the cloud point, no liquid-like layer
was formed at the attractive wall. However, for larger $n_0$ a
moderately thick layer was observed. The histogram of the interfacial
profile $\rho(\sigma,z)$ presented below correspond to the choice of
parameters $n_0=0.089$ $\tilde T=\tilde T_c=1.384$, with a wall strength
$\epsilon_w/k_BT=6.0$. The resulting liquid-like layer was found to
have a thickness of approximately $14\bar\sigma$. This 
is sufficiently large that the liquid-vapor interface can be considered
to be essentially decoupled from either wall, a view that was confirmed by
comparing the liquid and vapor properties on either side of the interface
with those obtained in the corresponding bulk simulations.

\begin{figure}[h]
\includegraphics[width=8.0cm,clip=true]{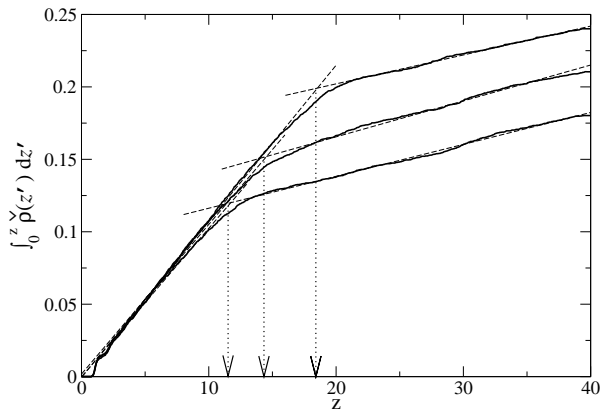}

\caption{Illustration of the procedure for locating the interface
center, as described in the text. Shown is the
running integral $\int_0^z\check\rho(\sigma,z^\prime)dz^\prime$ for
three example instantaneous density profiles that span the typical range of
fluctuations. Dashed lines denote linear fits to the running integral
in the pure liquid and vapor regions away from the interface. Dotted
arrows show the position of the interface center determined from the
intersection of the fits for each profile.}

\label{fig:average_sig}
\end{figure}

In the course of the simulations, the form of $\rho(\sigma,z)$ was
observed to exhibit large, slow fluctuations in which the layer
thickness varied in the range $12\bar\sigma$ to $18\bar\sigma$. These
fluctuations complicate the task of accumulating high statistics for
the intrinsic profile shape because an ensemble average over
independent configurations will be considerably smeared out in the $z$
direction. To circumvent this problem we have accumulated a {\em
centered profile} $\rho(\sigma,z^\star)$, whereby the instantaneous
profile $\check\rho(\sigma,z)$ is first shifted with respect to some
nominal origin (center) before being accumulated in the histogram. The
center of $\check\rho(\sigma,z)$ was itself determined by measuring the
running integral $\int_0^z\check\rho(\sigma,z^\prime)dz^\prime$. This
quantity manifests a smoothed discontinuity at the interface center, the
location of which can be estimated from the intersection of respective
linear fits in the pure vapor and liquid regions well away from the
interface, as illustrated in fig.~\ref{fig:average_sig}. We note that
our centering procedure differs to that adopted in some other studies
of liquid-vapor interfaces (see eg \cite{FISCHER80,ALEJANDRE}), where
the interface origin is defined via the instantaneous Gibbs dividing
surface.  Nevertheless, we do not expect differences in the precise
definition to have repercussions for the qualitative features of
results. Indeed, as a check of our procedure, we have investigated the
effect of disabling particle transfer MC moves (leaving operational
only particle displacement and resizing moves). Doing so greatly
suppresses the fluctuations in the interface thickness and yields an
average profile, the shape of which agrees to within error with that
obtained from the centering technique.

\section{Results}
\label{sec:results}

\begin{figure}[h]
\includegraphics[width=7.5cm,clip=true]{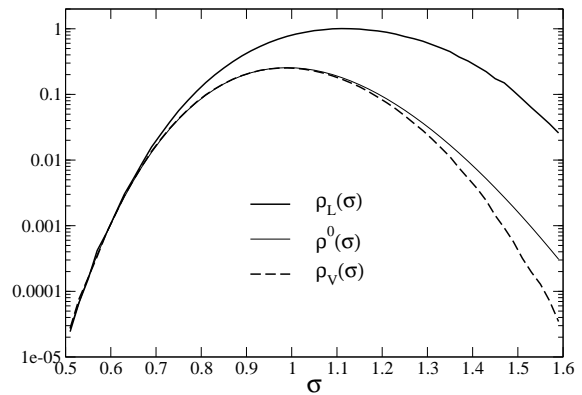}
\caption{Density distributions $\rho(\sigma)$ in the coexisting bulk phases at
$n_0=0.088$, $\tilde T=1.384$. Also shown is the parent form
$\rho^0(\sigma)$. Statistical errors are comparable with the line widths.}
\label{fig:rho_int}
\end{figure}

Before describing our findings for the liquid-vapor interfacial
properties, it is instructive to quantify the nature of the
fractionation in the pure phases away from the interface. This is done
in fig.~\ref{fig:rho_int}, which displays the single liquid and vapor
(daughter) phase distributions $\rho_L(\sigma)$ and $\rho_V(\sigma)$,
together with the Schulz parent distribution from which they derive,
all at the nominated bulk density of $n_0=0.089$. Clearly there
is a pronounced segregation of larger particles to the liquid phase.
Indeed, for $\sigma=\sigma_c$ the density in this phase exceeds
that of the vapor by a factor of two orders of magnitude--far greater 
than the ratio of the overall number densities in the coexisting phases
($\rho_L/\rho_V\approx 5$). One notes further that, as the particle size
decreases, the density distributions in both phases tend to the parent
form. The origin of this feature is traceable to the fact that the
smallest particles interact only very weakly (cf. eq.~\ref{eq:ljpot}),
and consequently their density is principally controlled by the imposed
chemical potential, rather than the number density.

\begin{figure}[h]
\includegraphics[width=9.5cm,clip=true]{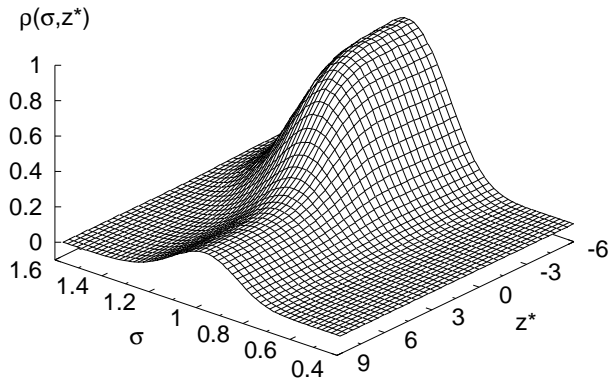}
\caption{Esimtates of the profile $\rho(\sigma,z^\star)$ near the liquid-gas
interface, determined by the procedure described in the text.}
\label{fig:rhosz}
 \end{figure}

\begin{figure}[h]
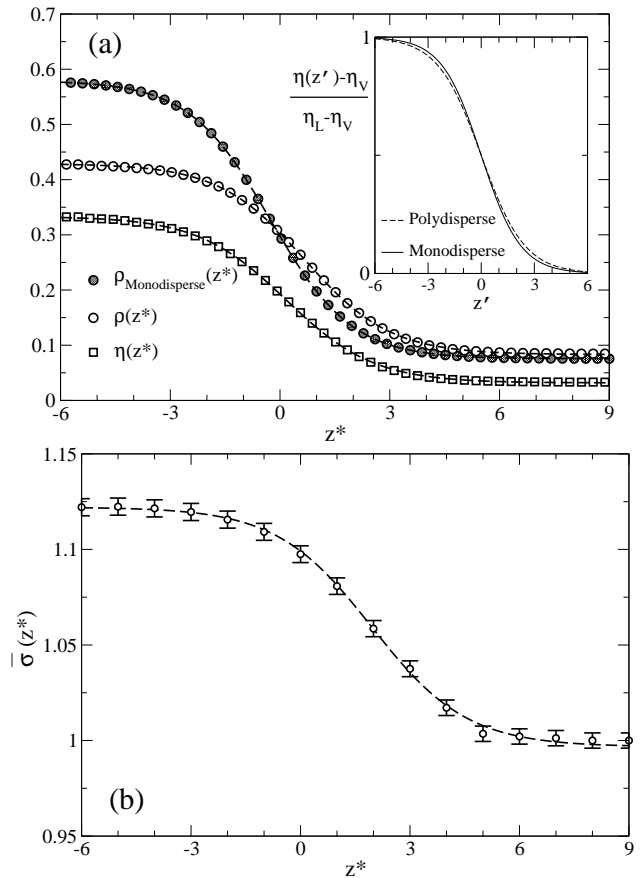

\includegraphics[width=7.8cm,clip=true]{fig6a.eps}
\includegraphics[width=8.3cm,clip=true]{fig6b.eps}

\caption{{\bf (a)} Estimates of the overall number density profile
$\rho(z^\star)$ and volume fraction profile $\eta(z^\star)$. Lines are
fits of the form eq.~\protect\ref{eq:tanh}. Also shown is the density
profile in the monodisperse limit at $\tilde T=1.080$, for which the
relative volume fraction fluctuations match those of the polydisperse
case (see text).  The inset shows a comparison of the volume fraction
profiles under these conditions. Statistical errors are smaller than
the symbol sizes. {\bf (b)} The variation across the interface of the
average local particle size, $\bar\sigma(z^\star)$. The line is a fit of
the form eq.~\protect\ref{eq:tanh}.}

\label{fig:avprofs}
\end{figure}

Turning now to the interfacial properties, fig.~\ref{fig:rhosz}
presents the measured form of the centered distribution
$\rho(\sigma,z^\star)$, accumulated as described in
sec.~\ref{sec:method}. The corresponding forms of $\rho(z^\star)$ and
$\eta(z^\star)$ (cf. eqs.~\ref{eq:denprof} and \ref{eq:vfracprof}) are
given in fig.~\ref{fig:avprofs}(a). Also shown
(fig.~\ref{fig:avprofs}(b)) is the variation of the average particle
size $\bar\sigma(z^\star)$ across the interface. 

We have attempted to fit $\rho(z^\star)$, $\eta(z^\star)$ and $\bar
\sigma(z)$ using a function of the standard $\tanh$ form
\cite{CAHN59,FISK69}:

\begin{equation}
\label{eq:tanh}
y(z^{\star}|\sigma)= y_a+y_d\tanh \left[ \frac {2(z^{\star}_{0}-z^{\star})}{\lambda}\right]\:,
\end{equation}
where $y_a=(y_L+y_V)/2$ and  $y_d=(y_L-y_V)/2$, with $y_L$ and $y_V$
the appropriate limiting pure phase quantity. Here $z^{\star}_{0}$
denotes the location of the interfacial mid-point, while $\lambda$ is a
measure of the interfacial half-width. It should be noted that the
fitting form eq.~\ref{eq:tanh} derives from mean field theories of
coexistence between {\em symmetrical} phases \cite{CAHN59,FISK69}, and
is not necessarily expected to hold for real asymmetric fluids which
lack particle-hole symmetry \cite{DEGENNES81,BARKER82,SZLEIFER90}.
Nevertheless eq.~\ref{eq:tanh} has been shown to provide a good
description of the liquid-gas interface density profile of the
monodisperse LJ fluid \cite{HOLCOMB93}, and indeed we do appear to
obtain good fits for all profiles, as evidenced in
fig.~\ref{fig:avprofs}. The associated mid-points and interfacial
widths are $z_0^\star=0.661$, $w\equiv2\lambda=9.3(2)$ for the
$\rho(z)$ profile, $z_0^\star=0.142$, $w=9.6(2)$ for $\eta(z)$, and
$z_0^\star=1.89$, $w=10.4(3)$ for $\bar\sigma (z)$.
Discrepancies in these estimates are presumably traceable to the
varying sensitivity of the respective observable to the interface
proximity. For example, because $\rho(z^\star)$ contains no direct
information on the variation of the concentration across the interface,
fits to its form possibly constitute a poorer estimator for the
interfacial width than $\bar\sigma(z^\star)$. We shall return to this point
below and in sec.~\ref{sec:concs}.

It is of interest to compare the interfacial properties of our
polydisperse fluid with those of the corresponding monodisperse system.
For reasons detailed in sec.~\ref{sec:concs}, we have chosen to perform
this comparison not at a given temperature and overall bulk number
density, but under two separate conditions relating to the {\em volume
fractions} of the coexisting pure phases. Firstly we consider, for the
monodisperse system, the temperature at which the value of the relative
fluctuation $2(\eta_L-\eta_V)/(\eta_L+\eta_V)$ equals that pertaining
to the polydisperse system at the state point under consideration
($\tilde T=1.384, n_0=0.089$). This occurs for $\tilde T=1.080$
\cite{WILDING95}. The associated monodisperse density profile is
included in fig.~\ref{fig:avprofs}(a), together with (in the inset) a
comparison of the volume fraction profiles of both systems. Fitting the
monodisperse volume fraction profile via the $\tanh$ form
(eq.~\ref{eq:tanh}), yields a width estimate $w=8.8(2)$, ie. some
$10\%$ smaller than found for the polydisperse profile $w=9.6(2)$. An
alternative scenario compares the poly- and mono-disperse systems under
conditions of equal volume fraction difference $\eta_L-\eta_V$ (which
occurs for the monodisperse system at $\tilde T=1.048$). Under such
conditions the difference in the width estimates falls to $5\%$. We
discuss these findings further in sec.~\ref{sec:concs}.

\begin{figure}[h]
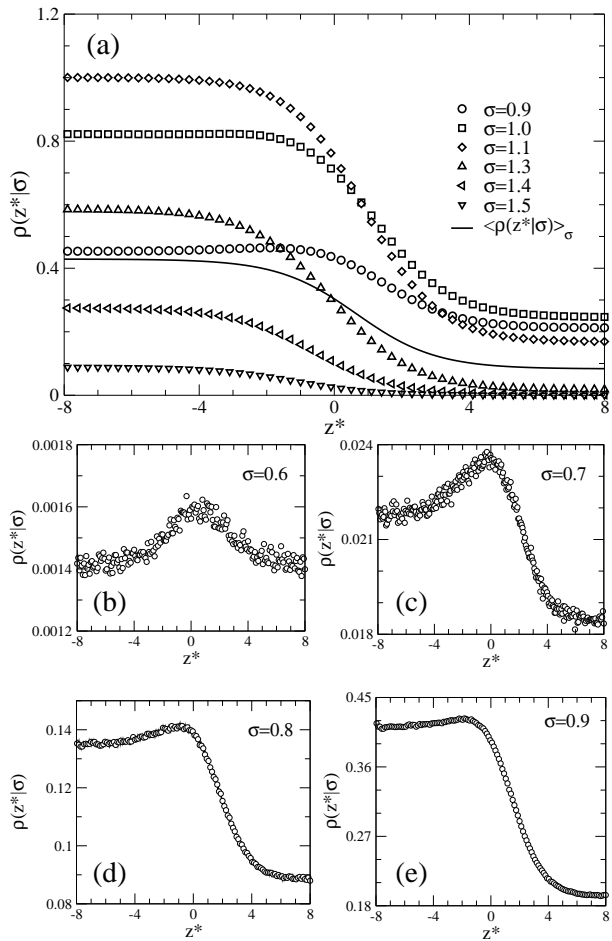

\includegraphics[width=8.0cm,clip=true]{fig7a.eps}
\includegraphics[width=8.0cm,clip=true]{fig7b.eps}

\caption{{\bf (a)} Estimates of interfacial density profiles
$\rho(z^\star|\sigma)$ for various $\sigma$ in the range $0.9\leq
\sigma \leq 1.5$. The solid line is the overall density profile
$\rho(z)$. Statistical errors are smaller than the symbol sizes. {\bf
(b)-(e)} show the corresponding profiles for a selection of $\sigma$ in
the range $0.6\leq \sigma \leq 0.9$. The statistical quality of the
data becomes poor for $\sigma\lesssim 0.6$ due to the low parent
density (cf. fig.~\protect\ref{fig:rho_int}). Statistical errors are
comparable with the spread of data points.}

\label{fig:prof_fits}
\end{figure}

Given knowledge of the distribution $\rho(\sigma,z^\star)$, it is
straightforward to extract the profiles for given individual species
i.e. the forms of $\rho(z^\star|\sigma)$. A representative selection of
these is presented in figs.~\ref{fig:prof_fits}, from which one
observes that for smaller values of $\sigma$, a pronounced
``segregation'' peak occurs in the density near the interface. It is
interesting to note that for the smallest particle size for which
reasonable statistics could be obtained ($\sigma=0.6$), the densities
on either side of the interface barely differ, in accord with the
coincidence (noted above) of the daughter distributions for small
$\sigma$. Notwithstanding this, a clear $10\%$ enhancement of the
density of this species occurs at the interface. That the profile
$\rho(z^\star|\sigma=0.6)$ couples to {\em changes} in the overall
density $\rho(z^\star)$ but not its absolute value provides an
indication that the segregation phenomenon is associated with the
surface tension of the interface. This point is discussed further in
sec.~\ref{sec:concs}.

\begin{figure}[h]
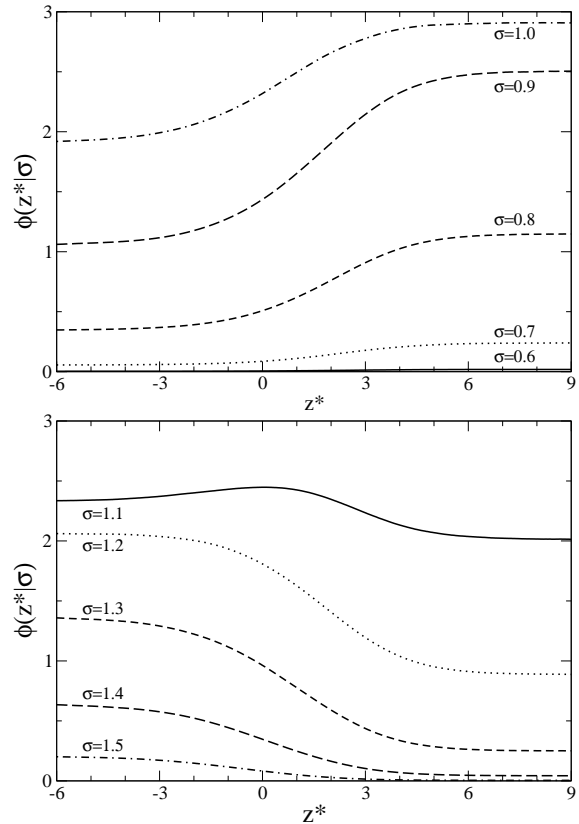

\includegraphics[width=7.5cm,clip=true]{fig8a.eps}
\includegraphics[width=7.5cm,clip=true]{fig8b.eps}
\caption{A selection of profiles of the local concentration
$\phi(\sigma,z^\star)$, as described in the text. Statistical errors do not
exceed the line widths.}
\label{fig:phis}
\end{figure}

The absolute height of the segregation peak is greatest for
$\sigma\approx 0.9-1.0$ and is apparent in all density profiles for
$\sigma \lesssim 1.0$. However, the associated segregation effect
actually extends to at least $\sigma =1.1$, as is evident from an
examination of plots of the local concentration $\phi(\sigma,z^\star)$
shown in fig.~\ref{fig:phis}. Specifically, the profile
$\phi(z^\star|\sigma=1.1)$ shows a clear peak at the interface. The fact that
a peak is visible in this profile, but neither in
$\rho(z^\star|\sigma=1.1)$ nor the concentration profiles for smaller
$\sigma$ values, can be explained as follows. For $\sigma<1.1$, the
concentration profiles are monotonically increasing with $z^\star$ and
the ratio of concentration in the vapor to that in the liquid increases
rapidly with decreasing $\sigma$.  For $\sigma>1.1$, however, the
profiles are monotonically decreasing and the ratio of concentration in
the vapor to that in the liquid decreases rapidly with increasing
$\sigma$. Accordingly, for $\sigma\approx 1.1$, the concentrations in the
two phases are closely matched. This matching enhances the visibility
of the segregation peak, which would otherwise be overwhelmed by the
large relative variation in the concentration profile across the
interface, as indeed occurs for smaller $\sigma$. This situation should
be contrasted with that for $\rho(z^\star|\sigma)$, where the ratio of
densities in the liquid and vapor become {\em more} closely matched as
$\sigma$ decreases (cf. fig.~\ref{fig:prof_fits}), thereby enhancing
the visibility of the segregation peak. Additionally, it appears likely
that any such peak is more distinguishable against a monotonically
decreasing profile as is the case for the density
$\rho(z^\star|\sigma)$, than for monotonically increasing ones as is
the case for the concentration profiles $\phi(z^\star|\sigma)$ for
$\sigma<1.1$

A more quantitative approach to the segregation effect is provided by a
quantity known as the symmetrized surface segregation, first introduced
in the context of binary liquid mixtures \cite{DAGAMA83,WINKELMANN01}.
This compares the interfacial density profiles of a nominated species
$\sigma$ with that of some reference species $\hat\sigma$: 

\begin{equation}
\Delta(z^\star|\sigma,\hat{\sigma})\equiv\frac{\rho(z^\star|\hat\sigma)-\rho_{L}(\hat\sigma)}{a_{\hat\sigma\sigma}}-\frac{\rho(z^\star|\sigma)-\rho_{L}(\sigma)}{a_{\sigma\hat\sigma}}\:,
\end{equation}
where the symmetrized concentrations are defined as

\begin{eqnarray}
a_{\hat\sigma\sigma} &=& \frac{\rho_L(\hat\sigma)-\rho_V(\hat\sigma)}{\rho_L(\sigma)-\rho_V(\sigma)+\rho_L(\hat{\sigma})-\rho_V(\hat{\sigma})}\;,\\
a_{\sigma\hat\sigma} &=& \frac{\rho_L(\sigma)-\rho_V(\sigma)}{\rho_L(\sigma)-\rho_V(\sigma)+\rho_L(\hat{\sigma})-\rho_V(\hat{\sigma})}\;.
\end{eqnarray}
Here $\rho_L(\sigma)$ and $\rho_V(\sigma)$ denote the density of a given
species in the bulk liquid and vapor phases respectively.

\begin{figure}[h]
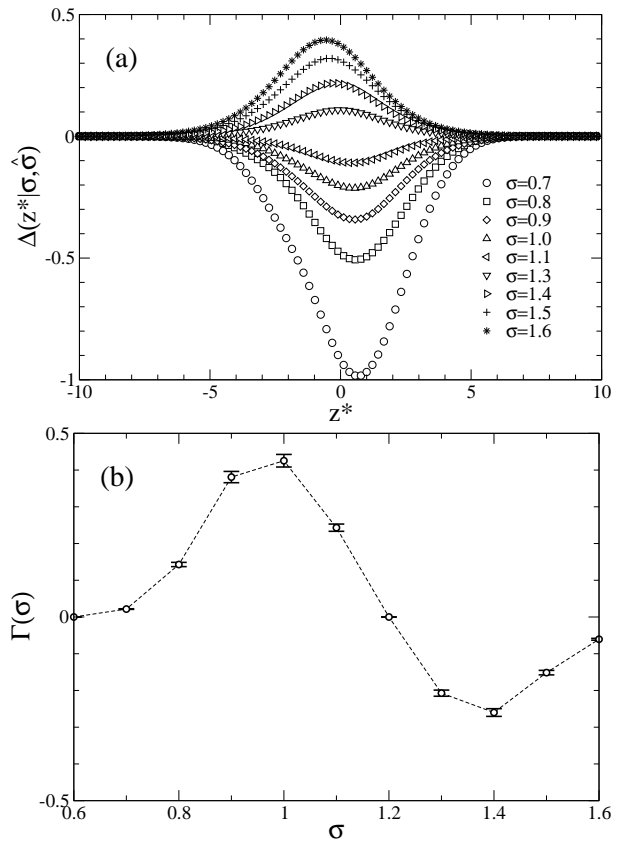

\includegraphics[width=8.0cm,clip=true]{fig9a.eps}
\includegraphics[width=8.0cm,clip=true]{fig9b.eps}
\caption{{\bf (a)} Estimates of the symmetrized surface segregation
$\Delta(z^\star|\sigma,\hat\sigma)$ (see text) for a choice of the
reference species $\hat\sigma=1.2$. Statistical errors are comparable
with the symbol size.
{\bf (b)}. The $\sigma$-dependent adsorption relative to
$\hat\sigma=1.2$.}
\label{fig:adsorb}
\end{figure}

$\Delta(z^\star|\sigma,\hat\sigma)$ is, by construction, zero in the two
pure phases, while at any point in the interfacial region it is either
negative or positive depending on the sign of the relative segregation
of the two species $\sigma$ and $\hat\sigma$. The relative adsorption
of species $\sigma$ with respect to species $\hat{\sigma}$ can then be
calculated as \cite{DAGAMA83}: 

\begin{equation}
\Gamma(\sigma|\hat\sigma)=-a_{\hat\sigma\sigma}\int_{-\infty}^{\infty}dz^\star~\Delta(z^\star|\sigma,\hat{\sigma}).
\label{eq:gamma}
\end{equation}
The density and concentration profiles discussed above show evidence of
surface segregation for $\sigma \lesssim 1.1$. In seeking to quantify
surface adsorption, it therefore seems reasonable to adopt as the reference
species $\hat\sigma=1.2$, for which no segregation was discernible. The
corresponding forms of $\Delta(z^\star|\sigma,\hat\sigma)$ are shown in
fig.~\ref{fig:adsorb}(a), together with the relative adsorption
$\Gamma(\sigma|\hat\sigma)$ in fig.~\ref{fig:adsorb}(b). The latter
figure clearly shows the adsorption peaking at around $\sigma=0.9-1.0$.
The population of particle whose size $\sigma>1.2$ are depleted in
comparison.  We note that the adsorption can be readily calculated via
alternative approaches, such as that based on the construction of
a Gibbs dividing surface for the reference species
\cite{ROWLINSON82,BELLIER02}. As a check on our procedure, we have also
applied this approach, finding results that are numerically practically
indistinguishable to those obtaining from eq.~\ref{eq:gamma}.

In a previous study of the liquid-vapor interface of a binary fluid
mixture \cite{WINKELMANN01}, the forms of $\Delta(z^\star)$ was argued
to provide a more reliable measure of the interfacial width than that
provided by $\tanh$ fits to the overall density profile $\rho(z^\star)$
alone, which tend to underestimate the true width. Adopting the
criterion \cite{WINKELMANN01} that the width is given by the range of
$z^\star$ for which $|\Delta(z^\star)|$ exceeds $5\%$ of its maximum
value, knowledge of $\Delta(z^\star|\sigma,\tilde\sigma)$ provides 
estimates of the interfacial width $w(\sigma)$ for each $\sigma$. This
quantity exhibits a spread of values, the largest corresponding to the
largest species, $w(\sigma=1.6)=10.4(3)$, a values which indeed clearly
exceeds that of $w=9.3(2)$ arising from the $\tanh$ fit to the density
profile $\rho(z^\star)$. It is, however, somewhat closer to that
obtained from the fit to $\eta(z^\star)$  (i.e. $w=9.6(2)$), and agrees well
with the estimate $w=10.4(2)$ obtained for the fit to
$\bar\sigma(z^\star)$. We postpone further discussion of this
comparison until sec.~\ref{sec:concs}.

\begin{figure}[h]
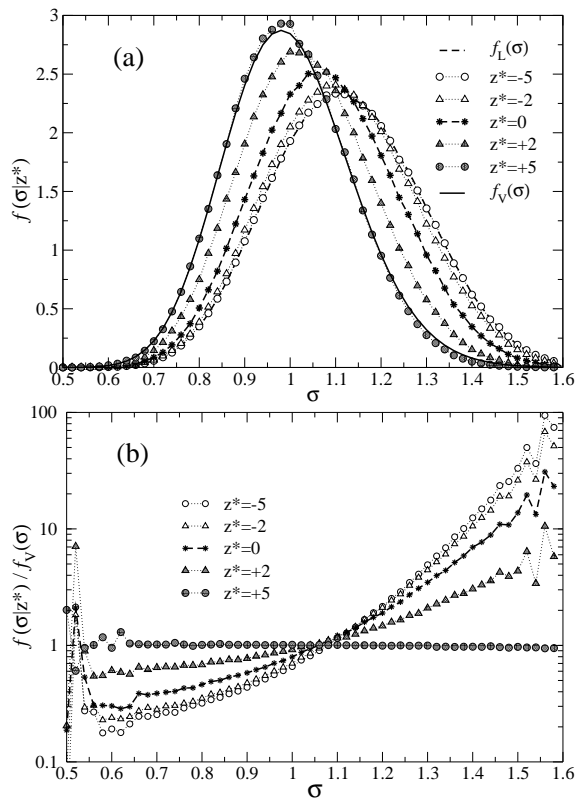

\includegraphics[width=7.5cm,clip=true]{fig10a.eps}
\includegraphics[width=7.5cm,clip=true]{fig10b.eps}
\caption{{\bf(a)} Variation of the normalized local size distribution
$f(\sigma|z^\star)$ across the interface, as described in the text. Lines through symbols are
merely guides to the eye. {\bf (b)} The same data referenced with respect
to the vapor phase distribution, and expressed on a log scale.}
\label{fig:sizedist}
\end{figure}

Having examined the density and concentration profiles and associated
adsorption, we turn finally to consider the variation of the normalized
size distribution $f(\sigma)$ across the interface,
eq.~\ref{eq:locszdist}. The forms of $f(\sigma|z^\star)$ for selected
values of $z^\star$ spanning the interface are shown in
fig.~\ref{fig:sizedist}(a). From the figure, one observes that on
traversing from the pure liquid to the pure vapor, the size
distribution steadily narrows, while its mean shifts to lower $\sigma$
(cf. figs.~\ref{fig:rho_int},~\ref{fig:avprofs}(b)). The scale of
the evolution of the fractionation across the interface becomes more
apparent when referenced with respect to one of the pure phases (we
have chosen the vapor) and plotted on a log scale, as shown in
fig.~\ref{fig:sizedist}(b). 

\section{Discussion and Conclusions}
\label{sec:concs} 

In summary, we have performed a detailed simulation study of the
liquid-vapor interface of a polydisperse fluid. The main feature of our
results is the finding of a preferential adsorption of smaller particle
at the interface and a corresponding depletion of larger ones. These
findings are in broad accord with those of a density functional theory
for a polydisperse van-der-Waals fluid \cite{BELLIER02}. We note
however that the segregation phenomenon is not peculiar to the case of
polydisperse mixtures. Analogous effects are well known to occur in the
context of binary fluid mixtures, where one component is often
preferentially absorbed at the liquid-gas boundary (see eg. refs.
\cite{CHAPELA77,DAGAMA83,ALEJANDRE,SMITH00,PRANGE01,LATSEVICH}). The
segregation occurs because coating the interface in the more volatile
(less strongly interacting) of the two species reduces the surface
tension, as can be shown from the Gibbs adsorption equation 
\cite{ROWLINSON82}. In view of this, one might expect that for the
polydisperse system the smallest species should be maximally adsorbed.
By contrast, we find that the adsorption is actually greatest for an
intermediate size (cf. fig.~\ref{fig:adsorb}). This finding can be
understood from the fact that for our interparticle potential
(eq.~\ref{eq:ljpot}) the smallest particles interact only weakly with
one another (and indeed become ideal in the limit $\sigma\to 0$).
Coating the interface in such very small particles, cannot therefore
screen the larger ones and reduce the surface tension. 

We have examined various quantities which provide a potential measure
of the liquid-vapor interfacial width. Good fits of the $\tanh$ form
(eq.~\ref{eq:tanh}) were achieved for the profiles of the overall
number density, volume fraction and local average particle size.
Additionally, a further measure, the $95\%$ width of the symmetrized
surface segregation profile,  $\Delta(z^\star|\sigma,\tilde\sigma)$,
was examined \cite{WINKELMANN01}. This latter quantity contains
explicit information on the density profiles of the individual species
$\rho(z^\star|\sigma)$, which, it seems, is important in obtaining
reliable estimates of the interfacial width. By contrast, the overall
profiles $\rho(z^\star)$ and $\eta(z^\star)$, integrate out some, or
all, of the $\sigma$ dependence of $\rho(\sigma,z^\star)$. They thus
suppress information about species whose individual profile width is
large, but which contribute relatively little weight to the overall
density. This may result in an underestimate of the true length scale
over which the system properties deviate from their bulk values, and
indeed the discrepancies we find in the interfacial width for the
various profiles (sec.~\ref{sec:results}) provide some evidence for
this. We note further that if one simply examines the overall density
profile, one may miss important features such as the segregation peak,
which is apparent in $\rho(z^\star|\sigma)$ for small $\sigma$, but not
in $\rho(z^\star)$.  

Turning now to the effects of polydispersity on the width of the
liquid-vapor interface, it is stated in ref.\cite{BELLIER02} that
polydispersity broadens the interface compared to the monodisperse
limit. However, our results highlight the need for care when attempting
to isolate the intrinsic effects of polydispersity on the interfacial
width from those arising indirectly as a result of
polydispersity-induced alterations to the bulk phase behaviour. Indeed,
the latter may well constitute the dominant factor. For example,
introducing polydispersity of the form described by eq.~\ref{eq:ljpot}
(which is similar to that considered in \cite{BELLIER02}) tends to
raise the critical temperature sharply with respect to the monodisperse
limit \cite{BELLIER00,WILDING05}. Thus on increasing the degree of
polydispersity at a {\em given} temperature, one can expect the
interface to become sharper since the system moves deeper into the two
phase region. Our results provide an extreme example of this.
Specifically the temperature at which we studied phase coexistence in
the polydisperse system, $\tilde T=1.384$ exceeds the corresponding
critical temperature in the monodisperse limit, namely $\tilde
T_c=1.1876$ \cite{WILDING95}. Thus at $\tilde T=1.384$ no interface
would occur at all in the monodisperse limit.

Another issue which complicates comparison of interfacial widths for
polydisperse and monodisperse system is the choice of parent
density $n_0$. Owing to fractionation and the resulting smearing of the
phase diagram (cf. figs.~\ref{fig:schem}), the value of $n_0$ represents
a crucial factor in determining the properties of the coexisting bulk
phases and hence the interface between them. Again the present work
provides a case in point: although we studied the system at the
critical temperature, true phase coexistence was nevertheless
observable because $n_0$ was set less than its critical value. Indeed
one can obtain a sharper or a broader interface simply by changing
$n_0$ at constant $T$, a situation which contrasts with that in the
monodisperse limit.

Thus we believe that when trying to isolate the intrinsic effect of
polydispersity on interfacial widths, one should endeavor to perform
the comparison at points in the phase diagram that are in some sense
equivalent in terms of their relative deviation from criticality.
Matching the bulk volume fraction fluctuations (in preference to the
densities which do not provide equivalent information) seems one
reasonable way to do this. Doing so (cf. inset of
fig.~\ref{fig:avprofs}(a)) shows that polydispersity may only broaden
the interface marginally ($5\%-10\%$). We caution, however, that our
study is far from comprehensive in this regard, and it is possible that
interfacial segregation and its effect on the surface tension and
interfacial width might become more significant at state points further
removed from criticality \cite{BELLIER02}. This would be an interesting
avenue for further computational study.

\acknowledgments

The authors thank R. Evans for a useful discussion and acknowledge
support of the EPSRC, grant number GR/S59208/01


\begin{thebibliography}{99}


\bibitem{SALACUSE} J.J. Salacuse and G. Stell, J. Chem. Phys. {\bf 77},
3714 (1982).

\bibitem{LARSON99} R.G. Larson, {\em The Structure and Rheology of
Complex fluids} (Oxford University Press, New York, 1999).

\bibitem{CHAIKIN00} P. Chaikin in {\em Soft and Fragile Matter}, M.E.
Cates and M.R. Evans (eds.), IOP publishing, London, 2000.

\bibitem{CHANDRASEKHAR92} S. Chandrasekhar, {\em Liquid Crystals}
(Cambridge University Press, Cambridge, 1992)

\bibitem{ELIAS97} H.-G. Elias, {\em An introduction to Polymer Science} (Wiley-VCH
Publishers, Weinheim, 1997)

\bibitem{SOLLICH02} P. Sollich, J. Phys. Condens. Matter {\bf 14}, R79 (2002).

\bibitem{RASCON03} C. Rasc\'{o}n and M.E. Cates, J. Chem. Phys. {\bf 118},
4312 (2003).

\bibitem{BELLIER00}  L. Bellier-Castella, H. Xu and M. Baus, J. Chem. Phys. {\bf 113}, 8337 (2000).

\bibitem{BRYK} P. Bryk, A. Patrykiejew, J. Reszko-Zygmunt, S.
Sokolowski, D. Henderson, J. Chem. Phys. {\bf 111}, 6047 (1999).

\bibitem{PAGONA00} I.~Pagonabarraga, M.E.~Cates, G.A.~Ackland, Phys. Rev. Lett. {\bf 84}, 911 (2000).

\bibitem{BUZZ04} M. Buzzacchi, I. Pagonabarraga and N.B. Wilding, J.
Chem Phys. {\bf 121}, 11362 (2004).

\bibitem{KEGEL04} R.P.A. Dullens and W.K. Kegel, Phys. Rev. Lett. {\bf 92},
195702 (2004).

\bibitem{PIZIO01} O. Pizio, A. Patrykiejew and S. Sokolowski, Mol.
Phys. {\bf 99}, 57 (2001). 

\bibitem{KIM01} S.-C.~Kim, J. Chem. Phys. {\bf 114}, 9593 (2001).

\bibitem{BELLIER02}  L. Bellier-Castella, H. Xu and M. Baus, Phys. Rev.
{\bf E65}, 021503 (2002); M. Baus,  L. Bellier-Castella and H. Xu, J.
Phys. Condens. Matter {\bf 14}, 9255 (2002).

\bibitem{WILDING02} N.B.~Wilding and P.~Sollich, J. Chem. Phys. {\bf 116}, 7716
(2002).

\bibitem{WILDING05} N.B.~Wilding,  M.~Fasolo  and P.~Sollich (unpublished).

\bibitem{WILDING03} N.B.~Wilding, J. Chem. Phys. {\bf 119}, 12163 (2003).

\bibitem{WILDING04A} N.B. Wilding and P. Sollich,
Europhys. Lett. {\bf 67}, 219 (2004).

\bibitem{WILDING04B} N.B. Wilding, M. Fasolo  and P. Sollich,
J. Chem. Phys. {\bf 121}, 6887 (2004).

\bibitem{FISCHER80} J. Fischer and M. Methfessel, Phys. Rev. {\bf A22},
2836 (1980).

\bibitem{ALEJANDRE} J. Alejandre, Y. Duda and S. Sokolowski, J. Chem. Phys.
{\bf 118}, 329 (2003).

\bibitem{CAHN59}   J.W.~Cahn and J.E.~Hilliard, J. Chem. Phys. {\bf 28},
258 (1958); {\bf 31}, 688 (1959).

\bibitem{FISK69} S.~Fisk and B.~Widom, J. Chem. Phys. 50, 3219 (1969).

\bibitem{DEGENNES81} P.G.~de~ Gennes, J Phys. Lett. (Paris), {\bf 42}, L377 (1981).

\bibitem{BARKER82} J.A.~Barker, J.R.~Henderson, J. Chem. Phys., {\bf 76}, 6303 (1982).

\bibitem{SZLEIFER90} I. Szleifer, J. Chem. Phys. {\bf 92}, 6940 (1990). 

\bibitem{HOLCOMB93} C.D. Holcomb, P. Clancy, J.A. Zollweg, Mol. Phys. {\bf 78}, 437 (1993).

\bibitem{WILDING95} N.B. Wilding, Phys. Rev. {\bf E52}, 602 (1995).

\bibitem{DAGAMA83} M.M. Telo da Gama and R. Evans, Mol. Phys. {\bf 48},
251 (1983).

\bibitem{WINKELMANN01} J. Winkelmann, J. Phys. Condens. Matter {\bf 13},
4739 (2001).

\bibitem{ROWLINSON82} J.S. Rowlinson and B. Widom, {\em Molecular Theory
of Capilliarity} (Clarendon, Oxford, 1982).


\bibitem{CHAPELA77} G.A. Chapela, G. Saville, S.M. Thompson and J.S. Rowlinson, Journal of the Chemical Society, Faraday Transactions 2,
73(7), 1133 (1977)


\bibitem{SMITH00} P. Smith, R.M. Lynden-Bell and W. Smith, Mol. Phys.
{\bf 98}, 255 (2000).

\bibitem{PRANGE01} W. Prange, T. Kurbjuhn, M. Tolan and W. Press, J. Phys. Condens. Matter {\bf 13},
4957 (2001).

\bibitem{LATSEVICH} S. Latsevich and F. Forstmann. J. Chem. Phys. {\bf
107}, 6925 (1997).

\end{thebibliography}
\end{document}